\documentclass[10pt,a4paper]{article}
\usepackage[utf8]{inputenc}
\usepackage[english]{babel}
\usepackage[T1]{fontenc}
\usepackage{amsmath}
\usepackage{amsfonts}
\usepackage{amssymb}
\usepackage{graphicx}
\usepackage{caption}
\usepackage[textwidth=4cm, textsize=footnotesize]{todonotes}
\usepackage{subcaption}
\usepackage{slashbox}
\begin{document}
\title{Technical report: Impact of evaluation metrics  and sampling on the comparison of machine learning methods for biodiversity indicators prediction }
\author{Geneviève Robin$^1$, Cathia Le Hasif$^2$\\

 \small 1: CNRS, Laboratoire de Mathématiques et Modélisation d'Évry, UEVE, France \\
 
  \small 2: Laboratoire de Mathématiques et Modélisation d'Évry, UEVE, France
}
\maketitle

\abstract{Machine learning (ML) approaches are used more and more widely in biodiversity monitoring. In particular, an important application is the problem of predicting biodiversity indicators such as species abundance, species occurrence or species richness, based on predictor sets containing, e.g., climatic and anthropogenic factors. Considering the impressive number of different ML methods available in the litterature and the pace at which they are being published, it is crucial to develop uniform evaluation procedures, to allow the production of sound and fair empirical studies. However, defining fair evaluation procedures is challenging: because well-documented, intrinsic properties of biodiversity indicators such as their zero-inflation and over-dispersion, it is not trivial to design good sampling schemes for cross-validation nor good evaluation metrics. Indeed, the classical Mean Squared Error (MSE) fails to capture subtle differences in the performance of different methods, particularly in terms of prediction of very small, or very large values (e.g., zero counts or large counts). In this report, we illustrate this phenomenon by comparing ten statistical and machine learning models on the task of predicting waterbirds abundance in the North-African area, based on geographical, meteorological and spatio-temporal factors. Our results highlight that differnte off-the-shelf evaluation metrics and cross-validation sampling approaches yield drastically different rankings of the metrics, and fail to capture interpretable conclusions.}

\section{Introduction}

In this section, we start by describing the objective and approach of this report; in brief, we aimed at illustrating the difficulty of designing fair comparison of ML methods in biodiversity applications. Then, we present the biodiversity case study we chose to illustrate our work; namely, the prediction of waterbirds abundance in North-Africa.

\subsection{Objectives of the report}

In previous work aiming to evaluate prediction methods for biodiversity data~\cite{lori}, authors of this report have faced the challenge of designing fair evaluation procedures, in order not to bias the conclusions, and thus provide sound and reliable recommendations for practitioners. Indeed, designing robust cross-validation (CV) evaluation metrics proved difficult, with different approaches yielding different results. Thus, the present report aims to illustrate this phenomenon on a real-life example and to discuss the particular issues which arise in evaluation experiments. To do so, we focus on a representative data set of waterbirds abundance monitoring in North-Africa, described in Section~\ref{MWN}, which we argue is quite representative of many biodiversity monitoring data sets. Then, we designed an experiment comparing 10 different ML prediction methods with different CV and evaluation metrics, and highlight the fact that results are very difficult to interpret globally, as different approaches yield different rankings. We thus argue for the development of uniform evaluation procedures, and discuss several possibilities at the end of the report.

\subsection{Illustration of Mediterranean Waterbirds Network (MWN) data}
\label{MWN}

As a case study, we use the updated MWN data set for North-Africa. This data set~\cite{SAYOUD201711}, monitors 16 waterbird species across 785 ecological sites located in the North-Africa (a region comprising Algeria, Egypt, Libya, Morocco and Tunisia) between 1990 and 2018. Each species is monitored yearly through synchronized counts, to survey the number of individuals present in each ecological site. Thus, for each species, a table of dimension 785x28 corresponding to the bird counts observed at each site for each year is available. In addition, researchers from the Tour du Valat institute have collected supplementary information via web scraping. Namely, geographical and environmental predictors (altitude, latitude, etc.) about the different sites, meteorological variables about each year (temperature and precipitation anomalies, etc.), as well as anthropogenic factors (agriculture level, economic index, etc.) are available. This data set is quite representative of modern biodiversity data collections, as it displays data heterogeneity (qualitative and quantitative predictors), missing values, complex count distributions (with zero-inflation and over-dispersion). Thus, we argue that this choice of case study is relevant, and the results obtained below can be seen as generalizable to other geographical or ecological contexts. 

To simplify the methodology, we have reduced our analysis to a single species, the Northern Shoveler, which is quite common and thus had the advantage of displaying fewer missing values than most other waterbirds in the data set. In addition, we restricted ourselves to years subsequent to 2010, and sites which had at least 5 observed values. In the end, we had about 2,200 observed counts for the Northern Shoveler, each one associated to a 18-dimensional predictor vector, used to predict the counts.

\section{Materials and Methods}
\label{sec:matmet}
In this section, we describe the 10 (classical) prediction models compared in this study. Then, we describe the experiment, whose goal is to compare the results of different evaluation procedures, where the varying parameters are the choice of sampling method for cross-validation (CV), and the choice of evaluation metric.

\subsection{Prediction methods}

 In this study, the aim is to train several different predictive models on a training MWN dataset, to evaluate the quality of the predictions, and thus to compare the models. We use 10 different prediction methods which to two main families: supervised learning methods, and imputation methods. In addition, one of the methods is a baseline, naive method that predicts the counts of the test dataset by the average of the counts in the train dataset. Qualitatively, supervised learning methods use only the predictors to predict the new response, while imputation methods use the observed counts as a bonus, learning a dependency structure between the count observations. Below, we provide short description of each method, along with appropriate references.
 
\paragraph{Generalized Linear Models (R package GLMnet) \cite{methode_glmnet}} 
The GLMnet method is based on a generalized linear model (in our case Poissonian), with the following log-linear model for a (count, predictor) couple $(Y_{ij}, X_{ij})$, where $i$ denotes the site and $j$ the sampling time:
$$ E(Y_{i,j}) = \exp(\mu_{i,j}),$$ \hspace{1cm} $$\text{with }\mu_{i,j} = \alpha_1X_{i,j}^{(1)} + \alpha_2X_{i,j}^{(2)} + ... + \alpha_{17}X_{i,j}^{(17)} + b.$$
Recall the Poisson log-likelihood: $\text{Loglike}_{Poisson} = \sum_{j=1}^{n_a}\sum_{i=1}^{n_{s_j}}\{Y_{i,j}\mu_{i,j} - \exp(\mu_{i,j}) + cste\}$.

In order to estimate the parameters $alpha_k$ and b, the R package \texttt{glmnet}~\cite{glmnet} minimizes the log-likelihood with a LASSO penalty. The following cost function is minimized:
$$ -n^{-1}\text{Loglike}_{Poisson} +  \lambda\|\alpha\|_1.$$

\paragraph{glmmTMB}
This method is also based on a GLM, but this time the distribution chosen is a negative binomial distribution, which is designed to handle over-dispersion.
We used the function \texttt{glmmTMB} of the R package of the same name~cite{glmmTMB} with the distribution parameter `nbinom1' consisting of the linear parameterization of a binomial distribution: $V = \mu*(1+\phi)$, with V the variance, $\phi$ the dispersion parameter = $\exp{\eta}$, $\eta$ the linear predictor coming from the dispersion model (negative binomial), and as link function the logarithm.

\paragraph{K-NN} The K nearest neighbors method is a supervised learning method used in classification but also in regression. We used the function \texttt{knn.reg} from the R package \texttt{FNN}~\cite{FNN} to obtain a regression model. The distance used in this algorithm is the Euclidean distance. A cross-validation was performed on the parameter \texttt{K} : the number of ``neighbor'' to take into account.

\paragraph{RandomForest}

Random forests are ensemble methods based on decision trees. For this method, we used the function \texttt{RandomForestRegressor} of the Python package \texttt{sklearn}. The R package \texttt{reticulate} allowed us to use python code in Rstudio as well as to access R objects in Python code and vice versa using the commands \texttt{r.} and \texttt{py\$}.
The best parameters \texttt{ntrees} and \texttt{mtry} were designated by cross validation using the Python function \texttt{GridSearchCV}.

\paragraph{Gradient Boosting}
The gradient boosting is also an ensemble method, which boosts the performances of weak models by adjusting the weights of individuals according to the results of the previous model. Thus, at each iteration, we have a better performing model than the previous one.

The gradient boosting method we use is the so-called `GBDT' method (Gradient Boosted Decision Trees), we used the function \texttt{gbm} of the package \texttt{gbm}~\cite{gbm}.
We specified that the model is Poissonian, so the loss function is the log-likelihood of a Poisson distribution.
The goal of the algorithm is to find the best step by minimizing the cost function. It is thus an optimization problem that the algorithm solves by gradient descent.
We have selected the best hyperparameters with a cross validation on \texttt{n.trees} and \texttt{interaction.depth}.

\paragraph{TRIM}
The Trim method builds log-linear models to predict counts, it is often used in this context with animal data.
We used the function \texttt{TRIM} from the R package \texttt{rtrim} by choosing the model 2 (`linear trend'). This model considers an effect of the sites and the same effect of the years on the counts according to the following model: $$\mu_{i,j} = \exp(\alpha_i + \beta*(j-1)),$$ with $\alpha_i$ the average of the counts of the site i, $\beta$ the factor of average growth which is the same for each site at each year.

\paragraph{LORI}

The function \texttt{lori} of the package \texttt{lori}~\cite{lori} uses an imputation method whose procedure resembles that of \texttt{glmnet}. The main difference is that, in this model, there are interaction parameters. An interaction is the fact that one covariate has an effect on the observations depending on one or more other covariates. This term is very well explained in the article `Main effects and interactions in mixed and incomplete data frames'~\cite{interaction}. By introducing these interaction parameters, we arrive at the following model to fit:
$$ E(Y_{i,j}) = \exp(\alpha_1X_{i,j}^{(1)} + \alpha_2X_{i,j}^{(2)} + ... + b + \theta_{i,j}),$$
with $\theta_{i,j}$ the interaction parameter between site $i$ and year $j$.  We always have: $$\mu_{i,j} = \alpha_1X_{i,j}^{(1)} + \alpha_2X_{i,j}^{(2)} + ... + b$$
The cost function is the same as that of \texttt{glmnet}, with a penalty on the interaction parameters: $$ -n^{-1}\text{Loglike}_{poisson} + \lambda_1|\alpha\|_1 + \lambda_2|\theta|_*,$$
$\|.\|_*$ being the nuclear norm.

\paragraph{MissForest}

The function \texttt{missForest} of the R package \texttt{missForest}~\cite{missForest} is a simple imputation method using random forests. At each iteration, all variables are fixed except one, and a random forest model is created to predict the missing data of a variable.

The algorithm for this method proceeds by first imputing the missing data with mean imputation. The variables are ordered in ascending order of the number of missing data they contain. The imputation of the data of a variable is done in 2 steps: a random forest model is trained, then the missing data are finally predicted by this model. The algorithm controls the performance of the method with the RMSE.
As for \texttt{RandomForest}, the hyperparameters \texttt{ntree} and \texttt{mtry} were chosen by cross-validation.

\paragraph{MICE}

The function \texttt{mice} of the R package \texttt{MICE}~\cite{mice} is a multiple imputation method, which thus allows to estimate the missing data but also the variability linked to the imputation. The process is essentially the same as \texttt{missForest}. Indeed, the method \texttt{mice} also proceeds by iteration: at each iteration, all the variables, except one, are fixed, they are going to be used as predictors, and an imputation method is chosen for the unfixed variable. The process is reiterated \texttt{m} times, parameter defined beforehand.

An essential parameter of this algorithm is the imputation method for each of the variables, we use a cross validation on this parameter among the following methods: \texttt{pmm} (predictive mean matching), \texttt{midastouch} (weighted predictive mean matching) and \texttt{norm} (Bayesian linear regression).

\subsection{Sampling methods for cross-validation}

The wide dispersion of the counts questionned the choice of sampling for the train and test data sets. Indeed, some methods were more accurate on small counts, and others on large counts. Note that, we define large counts as counts above the average value of the counts' distribution. We thus had a great variability according to the presence of large counts in the train or in the test.
In order to control the impact of large counts (over-dispersion) on the simulations, we implemented four different sampling functions:

\begin{itemize}
    \item A completely random sampling; 
This sampling function consists in taking randomly 80 \% of the data and thus to constitute a data set train, the 20 \% remaining constituting the data set test. This is done with the function \texttt{createDataPartition} of R.
    \item A stratified sampling; This sampling function consists of having 80\% of the large counts in the training dataset and the rest in the test dataset. The same principle is applied for the small counts. The process is as follows: we first classify the counts into 2 categories: above average (large) and below or equal to average (small), then we stratify so as to have the same proportion of large counts as small in the training and test datasets. The training dataset still represents 80 percent of the counts in the end.
    \item A balanced sampling; A balanced sample is created by first classifying the counts as performed for the stratified sampling, then subsampling in each of the 2 classes such that the resulting sample contains 50\% large counts. Next, a random partitioning is performed on our sample so that we have 80\% of the data in the training dataset.
    \item A sampling only of the small counts (the large counts are fixed in the training data).
\end{itemize}

\subsection{Evaluation metrics}

The fact that we have very dispersed counts did not allow us to naturally define a good metric for our objective. Indeed, in practice, we observed that some methods gave more accurate predictions for small counts, while others predicted large counts better. To try to make the evaluation objective,
we therefore chose three different metrics:

\begin{itemize}
    \item the root mean squared error (the RMSE, this metric gives equal weight to small and large accounts),
    \item the relative RMSE which gives an error renormalized by the size of the count (this metric gives more weight to small counts),
    \item the deviance which represents the deviation from the best Poissonian model.
\end{itemize}
\vspace{0.1cm}

\section{Results}

For each sampling method and each metric, we produced ten repetitions of the experiments, and display the results as boxplots.

\subsection{Comparison based on RMSE}

\begin{figure}[h]
    \begin{center}
        \begin{subfigure}{0.45\linewidth}
        \centering
            \includegraphics[width=\linewidth]{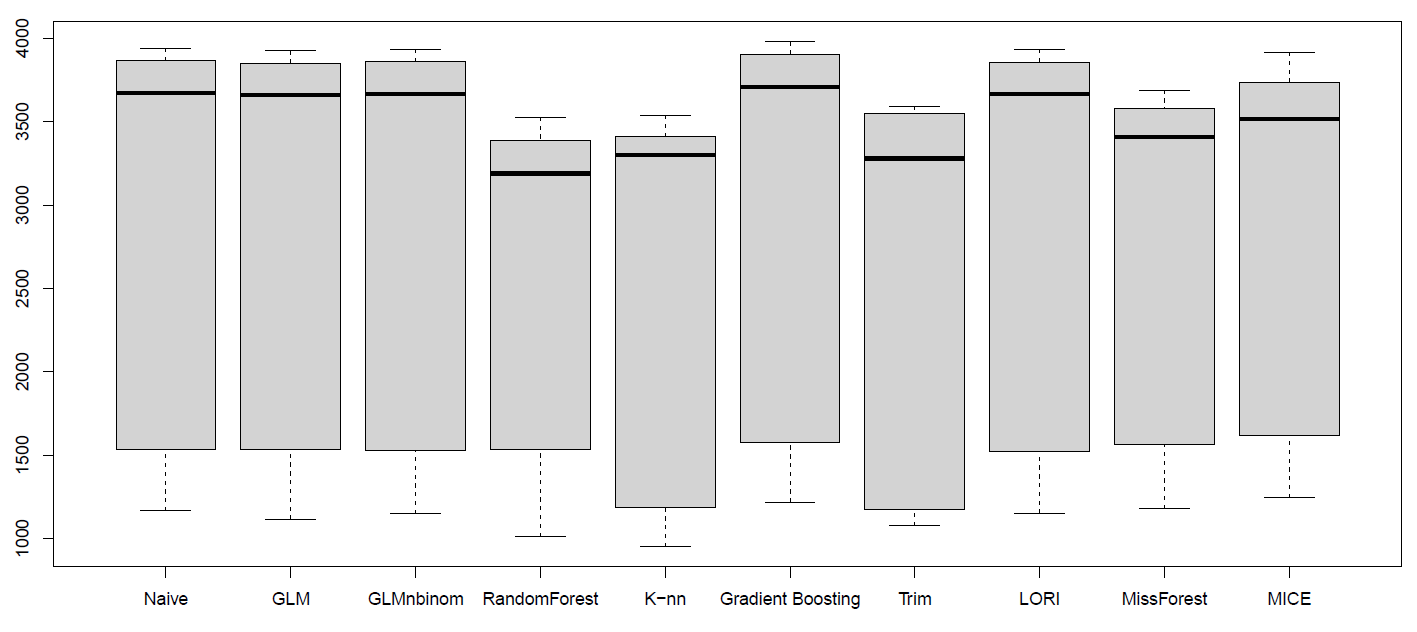}
            \caption{Completely random sampling}
            \label{subfig:rmse-random}
        \end{subfigure}
        \begin{subfigure}{0.45\linewidth}
        \centering
            \includegraphics[width=\linewidth]{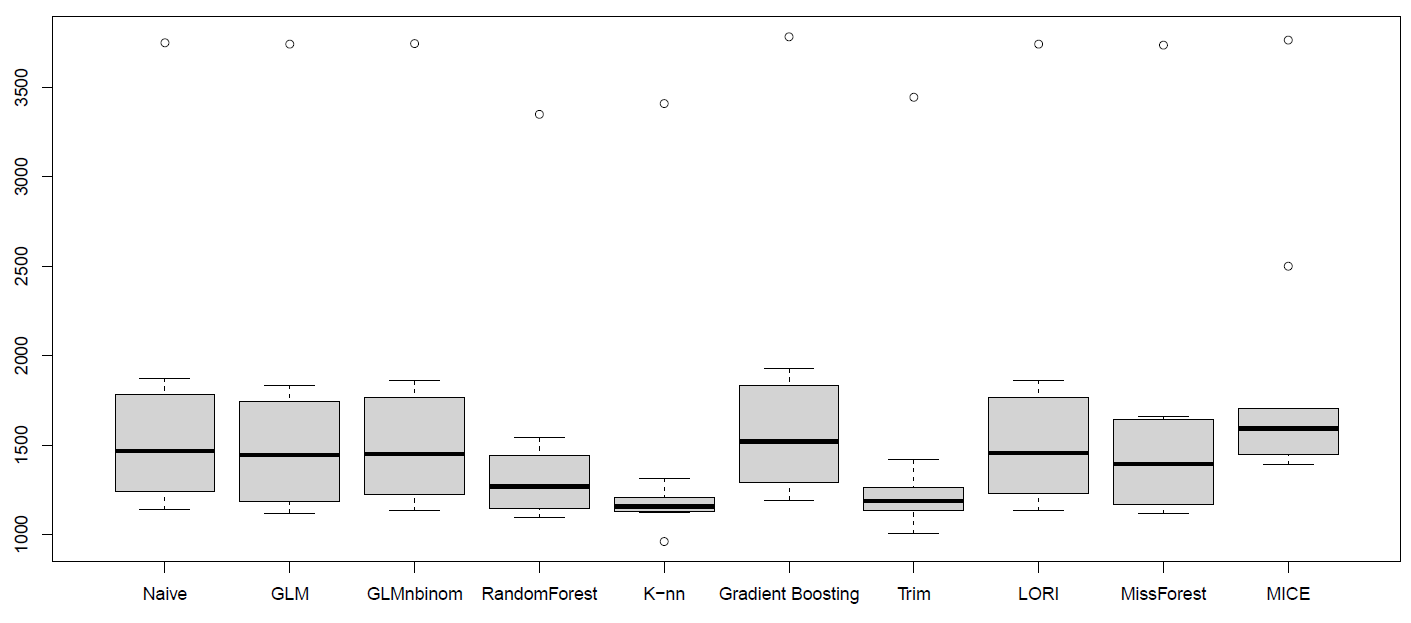}
            \caption{Stratified sampling}
             \label{subfig:rmse-strat}
        \end{subfigure}
        \begin{subfigure}{0.45\linewidth}
        \centering
            \includegraphics[width=\linewidth]{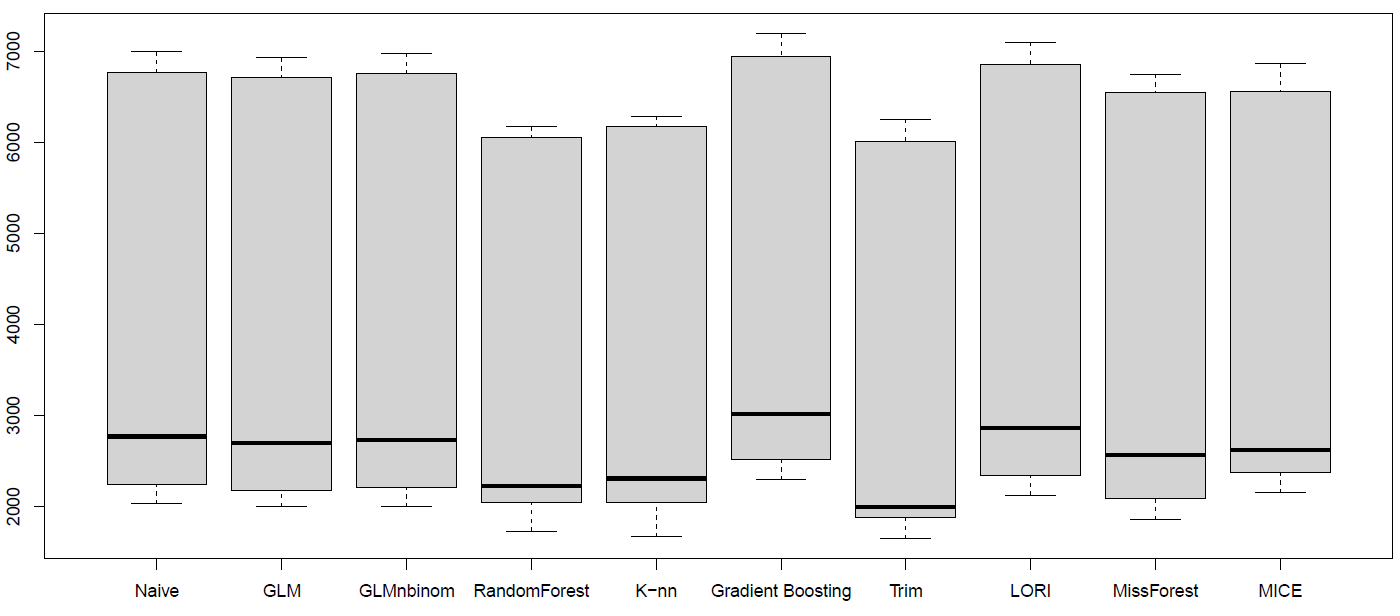}
            \caption{Balanced sampling}
             \label{subfig:rmse-balanced}
        \end{subfigure}
        \begin{subfigure}{0.45\linewidth}
        \centering
        \includegraphics[width=\linewidth]{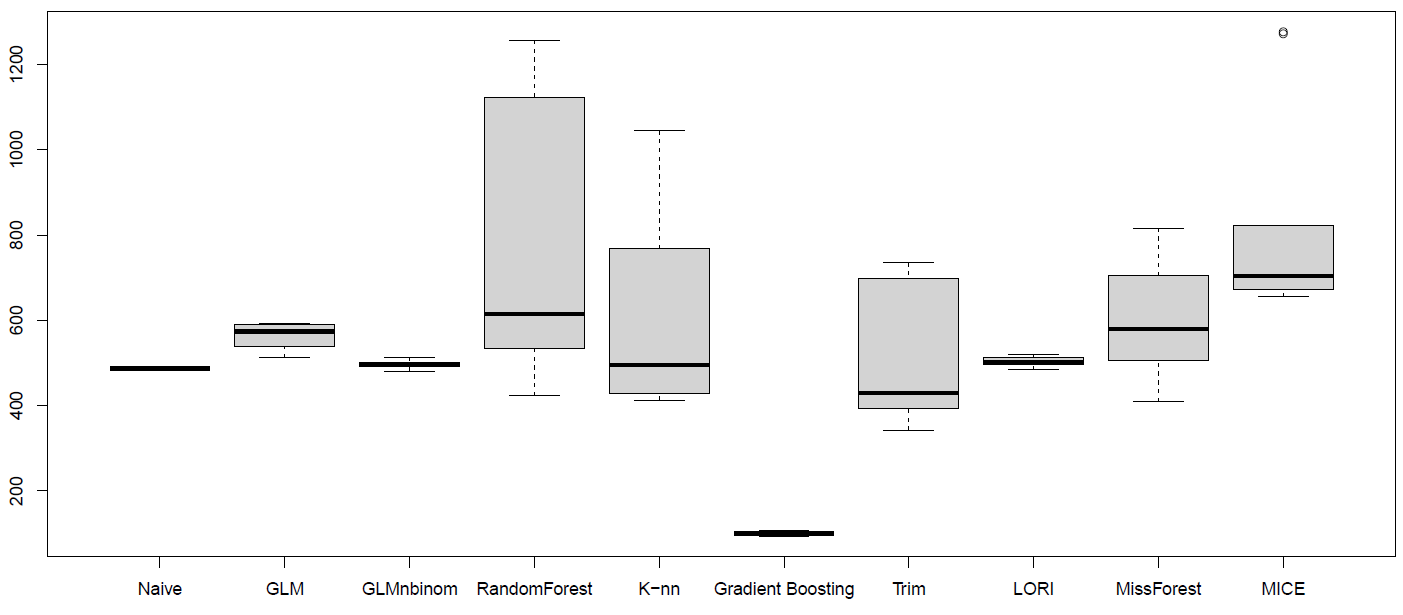}
        \caption{Sampling of small counts}
        \label{subfig:rmse-small}
        \end{subfigure}
    \end{center}
    \caption{Comparison of 10 prediction methods and 4 sampling approaches based on RMSE.}
    \label{erreurs_rmse}
\end{figure}

The first results, displayed in Figure~\ref{erreurs_rmse} present boxplots of the RMSE for the 10 compared methods, using the 4 different sampling methods described in Section~\ref{sec:matmet}. We observe that the classical approach for comparison, i.e. CV with completely random, stratified and balanced  sampling and evaluation with RMSE (see Figure~\ref{subfig:rmse-random}—\ref{subfig:rmse-balanced}) do not allow to discriminate the different methods which all display very large variability. After checking the 10 experiments individually, we observe that large errors are systematically due to the fact that large counts are sampled in the test set rather than the training set. In other words, large over-dispersion and zero-inflation make RMSE difficult to interpret. 

On the other hand, the obtained result based on sampling of small counts only allows to have smaller variability (see Figure~\ref{subfig:rmse-small}), and thus to compare the methods. However, this experiment is not representative of real-life problems, as large counts are systematically in the training set. From Figure~\ref{subfig:rmse-small}, we can only conclude that Gradient Boosting (GB) outperforms other methods in terms of prediction of the small counts.

\subsection{Comparison based on Relative MSE}

\begin{figure}[h]
    \begin{center}
        \begin{subfigure}{0.45\linewidth}
        \centering
            \includegraphics[width=\linewidth]{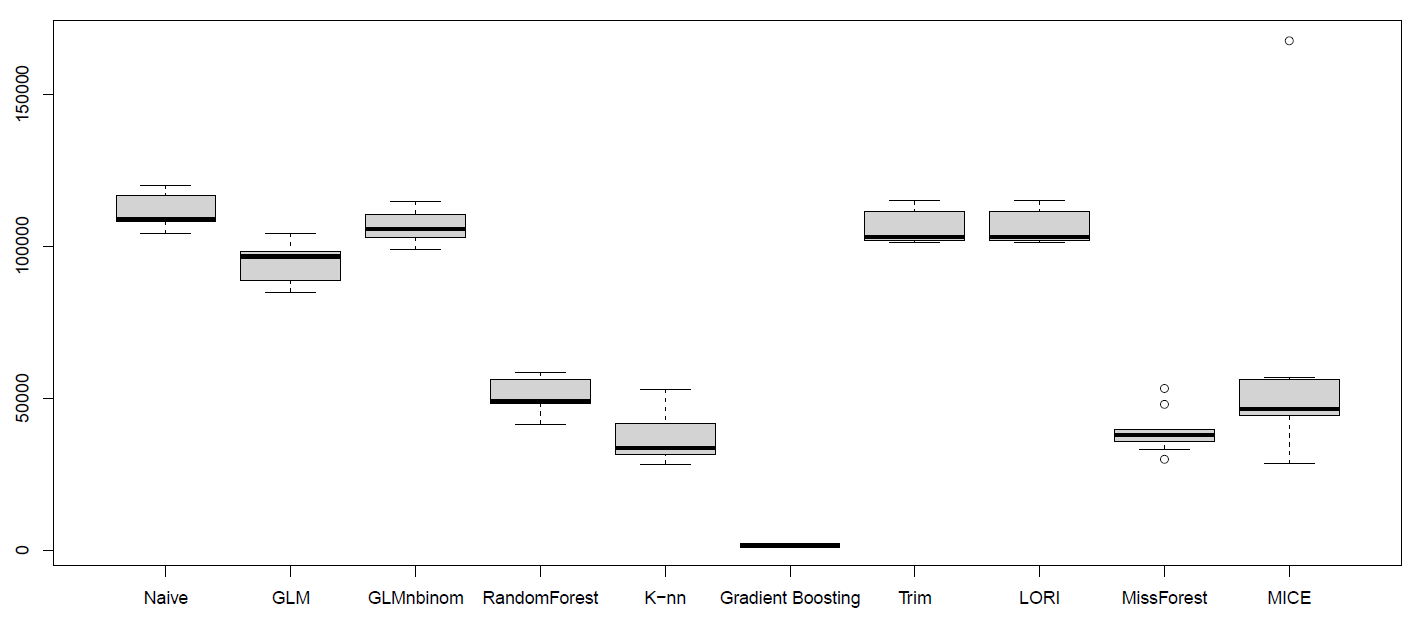}
            \caption{Completely random sampling}
            \label{relative_random}
        \end{subfigure}
        \begin{subfigure}{0.45\linewidth}
        \centering
            \includegraphics[width=\linewidth]{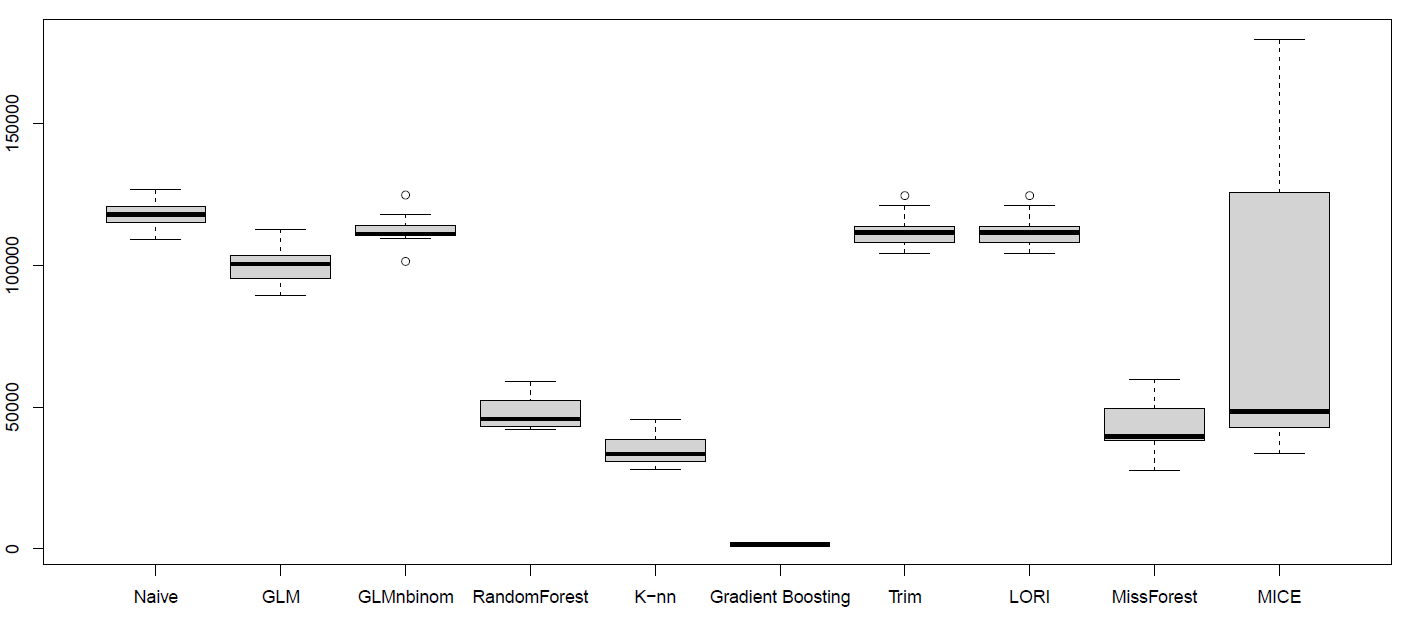}
            \caption{Stratified sampling}
            \label{relative_strat}
        \end{subfigure}
        \begin{subfigure}{0.45\linewidth}
        \centering
            \includegraphics[width=\linewidth]{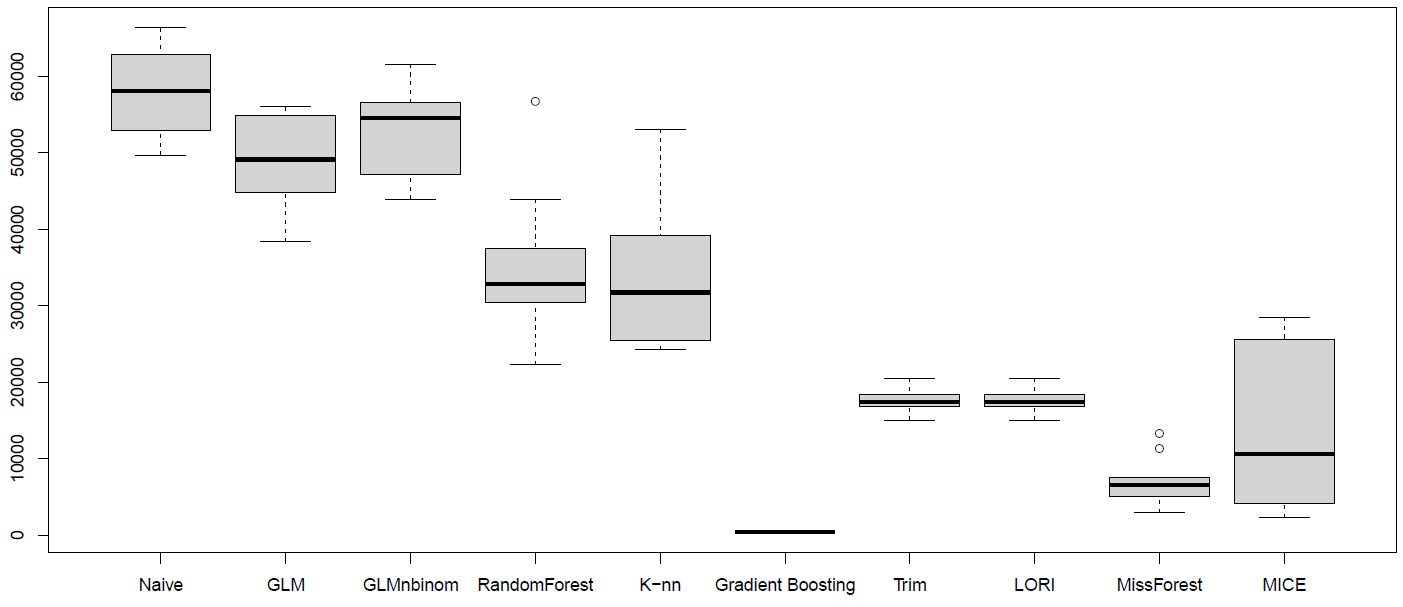}
            \caption{Balanced sampling}
            \label{relative_balanced}
        \end{subfigure}
        \begin{subfigure}{0.45\linewidth}
        \centering
        \includegraphics[width=\linewidth]{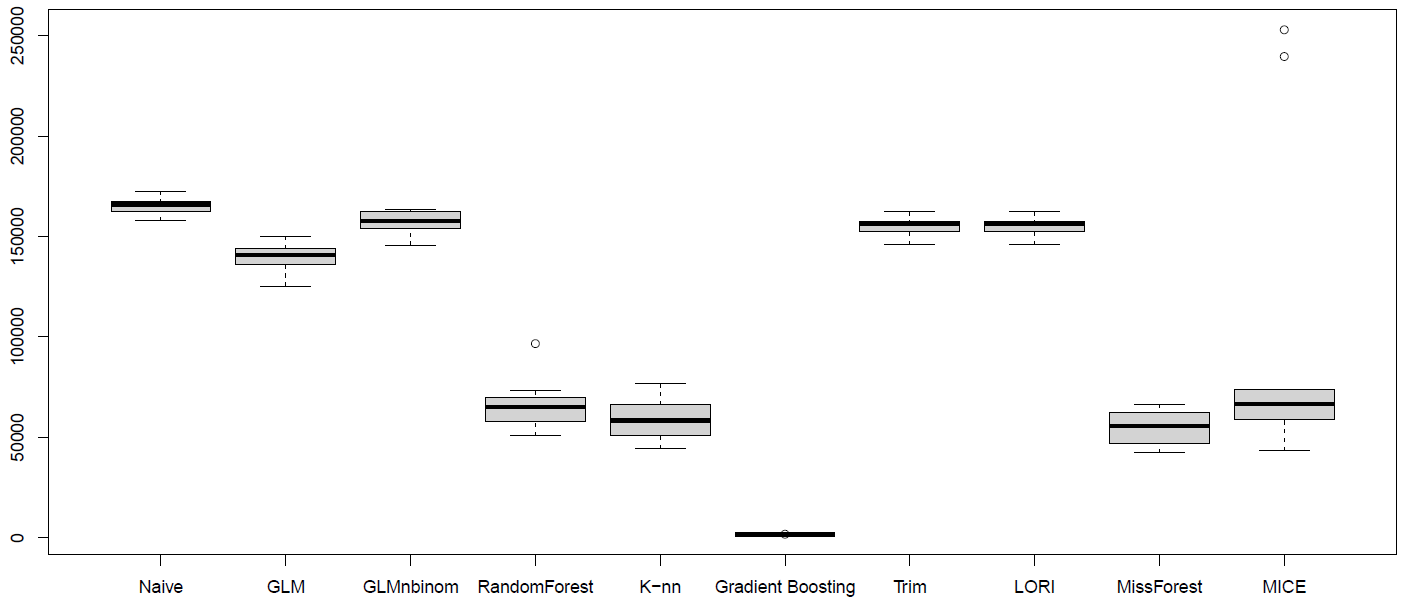}
        \caption{Sampling of small counts}
        \label{relative_small}
        \end{subfigure}
    \end{center}
        \caption{Comparison of 10 prediction methods and 4 sampling approaches based on Relative RMSE.}
\label{erreurs_relative}
\end{figure}

To try to alleviate the effect of counts dispersion, we display in Figure~\ref{erreurs_relative} boxplots of the Relative RMSE for the 10 compared methods, using the 4 different sampling methods described in Section~\ref{sec:matmet}. Note that, the Relative RMSE weights the RMSE by the size of the original count. Thus, the same absolute error is given more weight if it is made on a small count rather than a large count. We observe that the variability is much reduced compared to RMSE, and it is possible to compare the different methods. However, and important observation is that different sampling methods yields completely different rankings of the methods, and also different rankings compared to the experiment using RMSE.

\subsection{Comparison based on deviance}
\begin{figure}[h]
    \begin{center}
        \begin{subfigure}{0.45\linewidth}
        \centering
            \includegraphics[width=\linewidth]{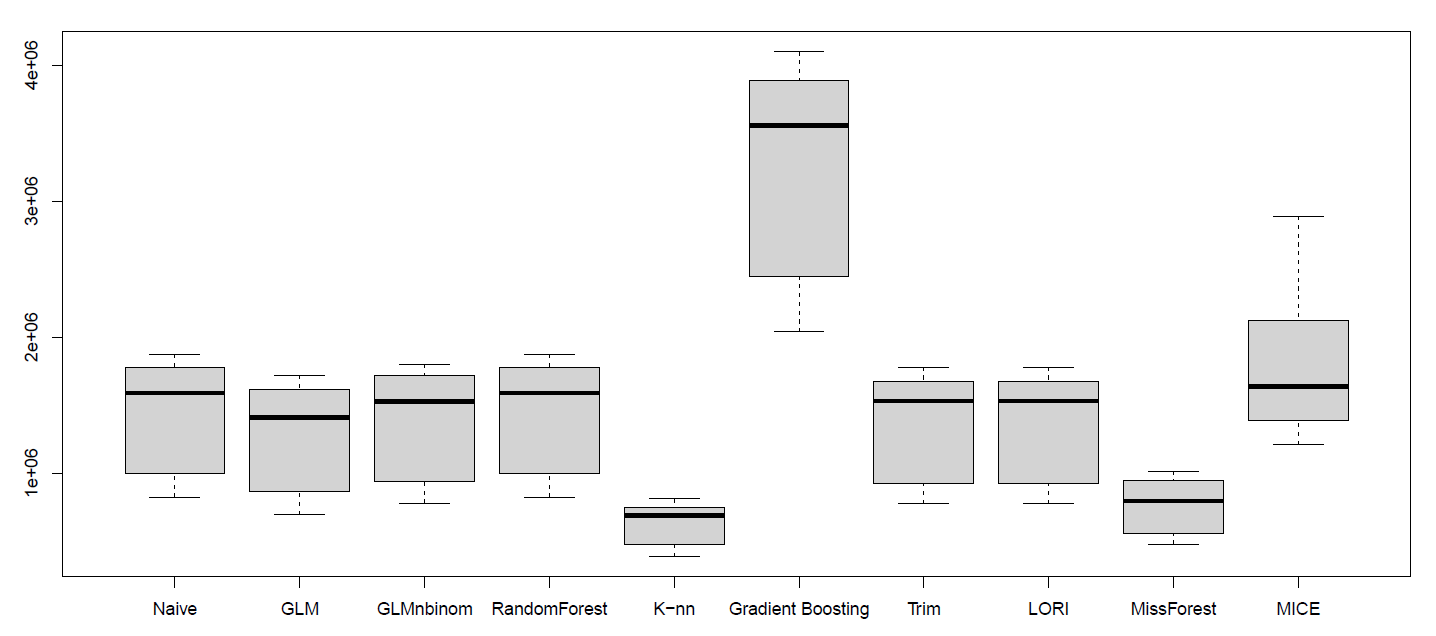}
            \caption{Completely random sampling}
        \end{subfigure}
        \begin{subfigure}{0.45\linewidth}
        \centering
            \includegraphics[width=\linewidth]{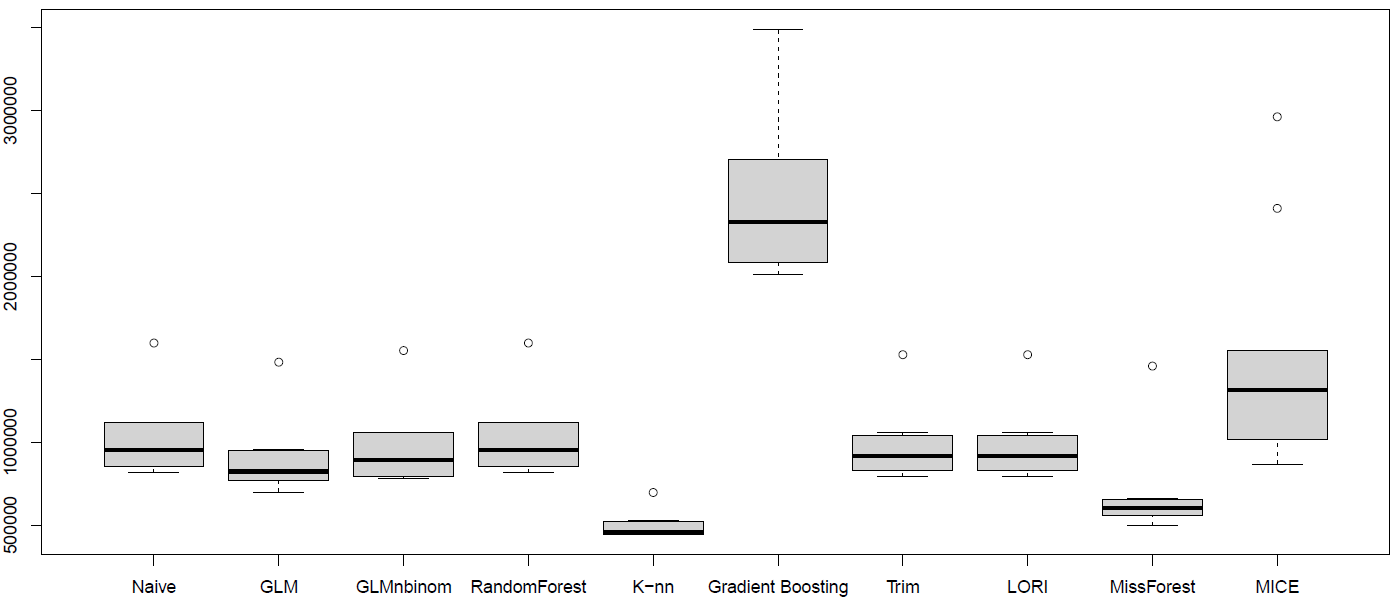}
            \caption{Stratified sampling}
        \end{subfigure}
        \begin{subfigure}{0.45\linewidth}
        \centering
            \includegraphics[width=\linewidth]{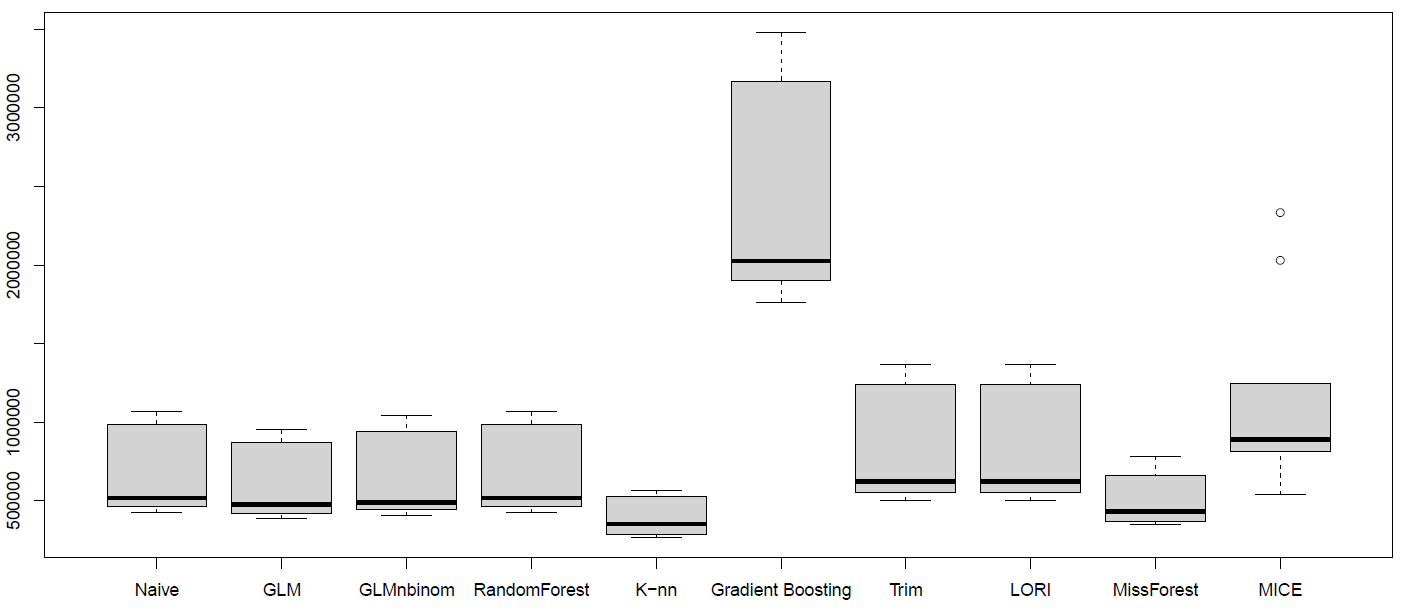}
            \caption{Balanced sampling}
        \end{subfigure}
        \begin{subfigure}{0.45\linewidth}
        \centering
        \includegraphics[width=\linewidth]{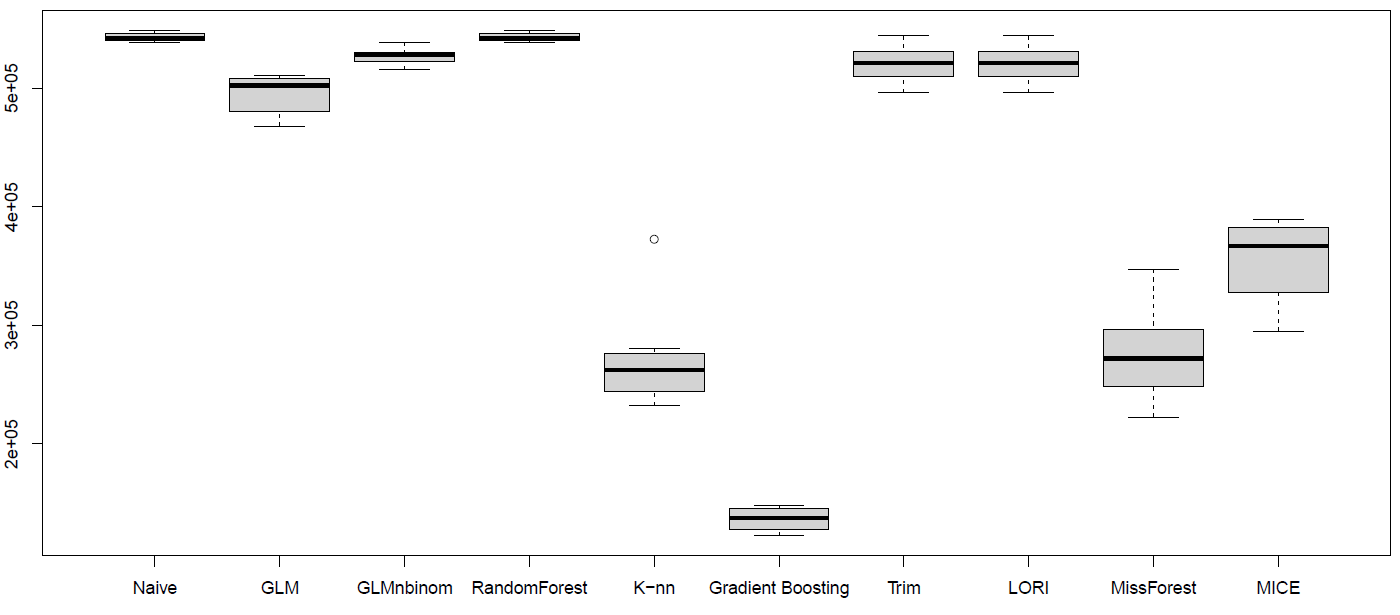}
        \caption{Sampling of small counts}
        \label{erreurs_random}
        \end{subfigure}
    \end{center}
        \caption{Comparison of 10 prediction methods and 4 sampling approaches based on Poisson deviance.}
\label{erreurs_deviance}
\end{figure}

Another way to alleviate the effect of counts dispersion,  is to use the Poisson deviance as an evaluation metric; results are displayed in Figure~\ref{erreurs_deviance}, with boxplots of the deviance for the 10 compared methods, using the 4 different sampling methods described in Section~\ref{sec:matmet}. We observe that, as in the Relative RMSE experiment, the variability is much reduced compared to RMSE, and it is possible to compare the different methods. However, the rankings are again very different across sampling methods, and compared to other metrics. In particular, we observe that GB rates very poorly compared to all other methods when evaluated with Poisson deviance, while it systematically outperformed all of them using the Relative RMSE.

\section{Discussion}

In this technical report, we illustrated the difficulty of comparing different ML methods for prediction of overdispersed, zero-inflated  species abundance data, on the example of the MWN data set. Table~\ref{top_3_methodes} presents a summary of the 3 best ranked methods for each evaluation metric and each sampling method.  It proved difficult to interpret the results globally because, depending on the metric, we do not obtain the same ranking of the methods, and despite the different sampling functions we observe a lot of variability, as shown in Table~\ref{top_3_methodes}.

\begin{table}[h]
\footnotesize
\centering
    \begin{tabular}{|l||*{4}{c|}}\hline
        \backslashbox[30mm]{Metric}{Sampling}
         & \makebox[10em]{Random} & \makebox[5em]{Stratified} & \makebox[5em]{Balanced} & \makebox[5em]{Small counts} \\\hline
        RMSE & 1. RandomForest & 1. K-nn & 1. Trim & 1. Gradient boosting
        \\ & 2. K-nn & 2. Trim & 2. Randomforest & 2. Trim
        \\ & 3.  Trim & 3. RandomForest & 3. K-nn & 3. K-nn
         \\\hline
        Relative RMSE & 1. Gradient boosting & 1. Gradient boosting & 1. Gradient boosting & 1. Gradient boosting
        \\ & 2. K-nn & 2. K-nn & 2. Missforest & 2. Missforest  
        \\ & 3. Missforest & 3.  RandomForest & 3. Lori & 3. K-nn
        \\\hline
        Deviance& 1. K-nn & 1. K-nn & 1. K-nn & 1. Gradient boosting
        \\ & 2. Missforest & 2. Missforest & 2. Missforest & 2. K-nn 
         \\ & 3. GLM & 3. GLM & 3. Randomforest & 3. Missforest
         \\\hline
    \end{tabular}
    \caption{Ranking of the 3 best methods for each evaluation metric and each sampling method.}
    \label{top_3_methodes}
\end{table}

Thus, we argue that it is necessary to define uniform evaluation procedures for the comparison of ML and, more generally, Artificial Intelligence (AI) prediction methods for biodiversity data. In particular, we argue for the development of more hybrid quantitative and qualitative metrics which would best capture the specificities of biodiversity data. For instance, we aim to investigate in future work the use of two-step evaluations, where methods are first ranked according to their ability to decipher small counts from large counts, and then on their prediction accuracy. Another possible approach would be to evaluate ML methods on their ability to recover known biological patterns.

\section{Acknowledgements}
The authors would like to thank all the organizations involved in the Mediterranean Waterbirds Network (MNW) who provided the waterbird data set used in this report: the GREPOM/BirdLife Morocco, the Direction générale des forêts (Algeria), the AAO/BirdLife Tunisia, the Libyan Society for Birds, the Egyptian Environment Affairs Agency, the Office national de la chasse et de la faune sauvage (ONCFS, France) and the Institut de recherche pour la conservation des zones humides méditerranéennes de la Tour du Valat (France). We are also grateful to all the field observers who participated in the North African IWC, thereby making this data set so rich, and to Laura Dami (Tour du Valat), Pierre Defos du Rau (Office Français de la Biodiversité) and Marie Suet (Tour du Valat) for their help.

\bibliographystyle{plain}

\bibliography{references.bib}

\end{document}